\documentclass[fleqn,usenatbib]{rasti}

\usepackage{newtxtext,newtxmath}

\usepackage[T1]{fontenc}

\DeclareRobustCommand{\VAN}[3]{#2}
\let\VANthebibliography\thebibliography
\def\thebibliography{\DeclareRobustCommand{\VAN}[3]{##3}\VANthebibliography}

\usepackage{subcaption}
\usepackage{graphicx}	
\usepackage{amsmath}	
\usepackage{xcolor}
\usepackage{soul}

\usepackage{array}      
\usepackage{booktabs}   
\usepackage{tabularx}   
\usepackage{ragged2e}   
\usepackage{makecell}   
\usepackage{ltablex} 

\title{e-CALLISTO FITS Analyzer: A Software Framework For CALLISTO Solar Radio Data}

\author[Liyanage et al.]{G. L. S. S. Liyanage,$^{1}$\thanks{E-mail: sahanslst@gmail.com, janaka@phys.cmb.ac.lk}
J. Adassuriya$^{1}$,
K. P. S. C. Jayaratne$^{1}$, 
C. Monstein$^{2}$
and P. K. Manoharan$^{3,4}$
\\
$^{1}$Astronomy and Space Science Unit, Department of Physics, University of Colombo, Sri Lanka\\
$^{2}$Istituto ricerche solari Aldo e Cele Daccò (IRSOL), Faculty of Informatics, Università della Svizzera italiana (USI), CH-6605 Locarno, Switzerland\\
$^{3}$Heliophysics Science Division, NASA Goddard Space Flight Center, Greenbelt, MD 20771, USA\\
$^{4}$The Catholic University of America, Washington, DC 20664, USA\\
}

\date{Accepted XXX. Received YYY; in original form ZZZ}

\pubyear{\the\year{}}

\begin{document}
\label{firstpage}
\pagerange{\pageref{firstpage}--\pageref{lastpage}}
\maketitle

\begin{abstract}
Solar radio bursts are signatures of dynamic processes in the solar corona, including particle acceleration and shock propagation associated with solar flares and coronal mass ejections. The e-CALLISTO archive, the largest near-real-time solar radio network with 150+ stations, produces large FITS volumes affected by interference, noise, and irregular frequency setups. Each 15-minute CALLISTO frame can split a burst across files, complicating event-level analysis. This work presents the e-CALLISTO FITS Analyzer, a unified, interactive, cross-platform application for processing e-CALLISTO dynamic spectra on Windows, macOS, and Linux. The application merges fragmented observations across time and frequency, applies an RFI-mitigation/background-subtraction pipeline with user-controlled clipping, and isolates bursts using an interactive polygon mask. It extracts a maximum-intensity backbone, supports outlier removal, and fits a power-law model to estimate drift rates, with uncertainties from the fit and backbone scatter. Shock height and speed are calculated using the Newkirk coronal density model, with optional $n$-fold density scaling. If the harmonic emission lane is used instead of the fundamental lane, its frequency drift is fitted separately. Analytical checks and empirical comparisons show that converting the harmonic lane to fundamental frequency changes estimated shock height and speed by only a few percent. For a Type II burst at Arecibo Observatory on 2 March 2022, the analyzer yielded an average drift rate of $-0.0400 \pm 0.0003$ MHz s$^{-1}$ and an average shock speed of $449 \pm 1$ km s$^{-1}$ at a height of $1.715 \pm 0.002~R_{\odot}$. It supports reproducible, event-focused SRB analysis and improves access to physically meaningful, uncertainty-quantified measurements.
\end{abstract}




\section{Introduction}

Solar radio bursts (SRBs) are critical diagnostic indicators of dynamic processes in the solar corona, providing insights into particle acceleration and shock wave propagation associated with solar flares and Coronal Mass Ejections (CMEs) \citep{bastian1998radio, raulin2005solar,gopalswamy2009coronal,white2024solar}. Among the various spectral classes, Type II bursts are of particular interest for space weather forecasting, as they are generated by magnetohydrodynamic (MHD) shocks often driven by CMEs, whereas Type III bursts trace the rapid travel of energetic electrons along open magnetic field lines \citep{cairns2010coherent, white2024solar,gopalswamy2008coronal}. These radio emissions, ranging from meter to decimeter wavelengths, serve as early warnings for geomagnetic storms and Solar Energetic Particle (SEP) events that can impact Earth's magnetosphere \citep{ameri2019properties, vourlidas2020radio, maia1999development}.

To monitor these transient events continuously, the Compound Astronomical Low-cost Low-frequency Instrument for Spectroscopy and Transportable Observatory (CALLISTO) network was established \citep{benz2005callisto,benz2009world}. Comprising more than 150 stations worldwide, this network utilizes low-cost programmable heterodyne receivers to observe dynamic solar radio events in the 45–870 MHz frequency range, generating spectra of a massive archive of Flexible Image Transport System (FITS) files \footnote{https://www.e-callisto.org/Data/data.html}\citep{benz2009world, bussons2023automatic}. Although this extensive coverage is invaluable, the sheer volume of data, thousands of files per station annually, presents a significant analysis bottleneck \citep{singh2019automated}. Raw spectra are frequently contaminated by terrestrial Radio Frequency Interference (RFI) and background noise, requiring robust filtering and processing techniques before meaningful physical parameters can be extracted \citep{hamidi2014mechanism}.

In recent years, various computational methods and software tools have been developed to address these challenges, though they often suffer from limitations in accessibility, accuracy, or functional scope. Statistical approaches, such as the method proposed by \citet{singh2019automated} using the Area Slope Index (ASI), attempt to distinguish bursts from noise automatically. However, these methods are often insensitive to weak bursts with low Signal-to-Noise Ratios (SNR < 5$\sigma$) and lack the capability to classify specific burst types accurately \citep{singh2019automated}. Similarly, the "Burst-Finder" algorithm developed by \citet{afandi2020burst} utilizes thresholding on resized spectra to detect bursts. While achieving a high detection rate, such fully automated systems are prone to false positives caused by RFI and are designed primarily for event logging rather than detailed physical analysis \citep{afandi2020burst}.

More recently, deep learning approaches have been applied to e-CALLISTO data. For instance, \citet{bussons2023automatic} introduced the "deARCE" method based on Convolutional Neural Networks (CNNs) like AlexNet. While these "black box" models are highly effective for binary classification (burst vs. no-burst) and generating event reports, they require massive training datasets and do not provide the interactive environment necessary for researchers to extract precise kinematic properties of the shock waves \citep{bussons2023automatic}. On the other hand, libraries such as 'pyCallisto' offer Python-based functions for background subtraction and flux integration \citep{pawase2020pycallisto}. However, 'pyCallisto' is a code library rather than a standalone application, requiring users to possess programming proficiency to script their own analysis pipelines, which creates a barrier for observational astronomers who require ready-to-use tools \citep{pawase2020pycallisto}.

Attempts to bridge the gap between automation and interactivity include the "Solar Radio Burst Analyzer" developed by \citet{hettiarachchi2024analysis}. This Python-based Graphical User Interface (GUI) assists in calculating drift rates using an "Intensity Matrix" method. Despite its utility, the software has notable limitations. First, it is semi-automated, relying heavily on the user to manually define start and end coordinates to isolate burst regions \citep{hettiarachchi2024analysis}. Second, it imposes rigid mathematical models on the data, assuming a linear fit for Type III bursts and an exponential decay model ($f = ae^{-(b+t)/t} + c$) for Type II bursts \citep{hettiarachchi2024analysis}. These assumptions may fail to capture the complex, non-linear morphological variations often observed in dynamic solar eruptions. Furthermore, e-CALLISTO observations are segmented into 15-minute files; long-duration events often span multiple files, yet existing tools often lack a seamless workflow to merge these fragmented segments into a continuous spectrum for uninterrupted analysis \citep{benz2009world, pawase2020pycallisto}. Due to the non-availability of the comprehensive analysis tool, some of the studies were limited to a few data files \citep{ansor2020effectiveness, wijesekera2018analysis} despite the vast archive of e-CALLISTO. Furthermore, the e-CALLISTO archive consists of data from different locations, which may not be identical in frequency span due to their ideal observation window within 45 - 870 MHz. Therefore, it is not easy to combine the fragmented radio bursts in two consecutive data files for the analysis \citet{giersch2017solar}.

To address these deficiencies, we present the "e-CALLISTO FITS Analyzer." Unlike the rigid modeling of \citet{hettiarachchi2024analysis} or the code-heavy nature of `pyCallisto` \citep{pawase2020pycallisto}, this new tool offers a flexible, unified GUI environment. It streamlines the ingestion of fragmented FITS files, automates file merging across time, and implements robust background subtraction to mitigate RFI. By providing interactive tools that allow for flexible tracking of burst backbones without restricting the user to specific mathematical shapes, the software facilitates the precise extraction of frequency drift rates and shock parameters. This tool aims to democratize access to rigorous solar radio analysis, enabling broader participation in space weather research.

\subsection{Strengths and novelty of the proposed framework}
\label{subsec:novelty}

Several software tools and scripts are already available for CALLISTO and
solar radio spectrogram analysis, and the e-CALLISTO FITS Analyzer is not
intended to replace them. These existing resources fall broadly into three
groups, compared in Table~\ref{tab:software_comparison} across the parameters
most relevant to an observational user. Archive and visualisation
utilities including the quicklook products of the e-CALLISTO archive and the
IDL and MATLAB plotting routines distributed on the e-CALLISTO software
page which allow FITS spectra to be read and displayed, but are oriented toward
inspection rather than interactive parameter extraction. Scriptable Python
libraries such as \texttt{pyCallisto} \citep{pawase2020pycallisto} and
\texttt{radiospectra} \citep{radiospectra} provide programmatic operations,
including time- and (for \texttt{radiospectra}) frequency-axis combination and
background subtraction, but require programming proficiency and stop short of
shock-parameter estimation. Automated and GUI-based burst tools, including
machine-learning detectors and earlier graphical drift-rate utilities, target
burst detection or selected measurements, the former trading interactivity for
throughput and the latter often imposing fixed analytical fit models. Against
this background, the purpose of the present work is to provide an integrated,
reproducible, and user-accessible workflow for event-level analysis of
e-CALLISTO FITS observations.

The novelty of the framework lies in combining, within a single cross-platform
graphical environment, a set of operations that are usually handled
separately: archive-based data retrieval, FITS parsing, time-axis merging,
frequency-axis merging, RFI mitigation and background suppression, interactive
burst isolation, maximum-intensity backbone extraction, outlier removal,
drift-rate fitting, and shock-parameter estimation. As
Table~\ref{tab:software_comparison} indicates, the distinguishing
characteristic of the e-CALLISTO FITS Analyzer is therefore not any individual
operation, several of which are also available in existing libraries, but
their integration into one transparent, open-source pipeline that runs
interactively on Windows, macOS, and Linux without programming. In particular,
the deterministic RFI-mitigation stage, which performs automated hot-channel
detection and repair before background subtraction, is not part of the standard
e-CALLISTO processing chain and represents one of the framework's principal
additions.

\begin{table*}
\centering
\caption{Comparison of the principal software approaches used for CALLISTO / e-CALLISTO
dynamic-spectrum analysis, across the parameters most relevant to an observational user:
interface model, cross-platform availability, open-source status, flexibility, and computational
load. The present work is distinguished not by any single operation, but by
combining an interactive cross-platform GUI with event-level time- and frequency-axis merging,
RFI mitigation, and Type~II shock-parameter estimation in one open-source workflow.}
\label{tab:software_comparison}
\footnotesize
\renewcommand{\arraystretch}{1.3}
\begin{tabular}{>{\raggedright\arraybackslash}p{3.3cm}
                >{\raggedright\arraybackslash}p{2.3cm}
                >{\raggedright\arraybackslash}p{2.6cm}
                >{\raggedright\arraybackslash}p{1.9cm}
                >{\raggedright\arraybackslash}p{1.7cm}
                >{\raggedright\arraybackslash}p{2.5cm}}
\toprule
Tool or approach & User interface & Cross-platform (Win/\allowbreak macOS/\allowbreak Linux) & Open source & Flexibility & Time to run \\
\midrule
e-CALLISTO archive / quicklook utilities & Web service & Web-based & --- & Low & Instant (server-side) \\
CALLISTO IDL / MATLAB plotting routines & Script (IDL/MATLAB) & via runtime & Partial & Low & Fast (plotting) \\
pyCallisto & Library (Python) & Yes & Yes & Moderate & Fast (scripted) \\
radiospectra (SunPy-affiliated) & Library (Python) & Yes & Yes & Moderate & Fast (scripted) \\
Automated burst detection (e.g.\ deARCE) & Script / ML pipeline & Yes & --- & Low & Heavy (model training) \\
Solar Radio Burst Analyzer & GUI (desktop) & Yes (Python) & --- & Moderate & Interactive (user-paced) \\
\textbf{e-CALLISTO FITS Analyzer (this work)} & \textbf{GUI (desktop)} & \textbf{Yes} & \textbf{Yes} & \textbf{High} & \textbf{Interactive (user-paced)} \\
\bottomrule
\multicolumn{6}{@{}p{15cm}@{}}{\smallskip\footnotesize\textit{Notes.}
\emph{Flexibility} grades the degree of user control over the processing chain and the breadth of
supported operations: low = fixed products or a single imposed model; moderate = full scripted or
guided control, but limited analysis scope or imposed fit models; high = interactive control over
burst isolation, fitting, and lane selection without an imposed burst geometry.
\emph{Time to run} is a qualitative indication of computational load for typical event-level
processing, not a benchmarked wall-clock time; the interactive tools are user-paced rather than
batch, so their effective runtime is dominated by user interaction rather than computation.
``via runtime'' = a proprietary IDL/MATLAB environment is required; ``Partial'' (open source) =
source is freely available but requires that proprietary runtime; ``---'' = not stated in the
cited source.} \\
\end{tabular}
\end{table*}

A key strength of the framework is its treatment of fragmented e-CALLISTO
observations. Since routine CALLISTO data are stored as short observing frames
and may also differ in usable frequency coverage between stations or receiver
configurations, long-duration solar radio bursts can be difficult to analyse as
continuous events. The software addresses this by supporting both time-domain
and frequency-domain combination of compatible FITS files before quantitative
analysis, so that a burst distributed across separate files or frequency
sub-bands can be inspected and measured as a single continuous dynamic
spectrum. This reduces manual preprocessing and improves the consistency of
subsequent measurements.

The second strength is the balance between automation and user control. Fully
automated burst-detection methods are valuable for event logging and catalogue
construction, but they can be sensitive to radio-frequency interference, weak
emission, or complex burst morphology, while purely script-based workflows
require programming experience and make repeated visual inspection cumbersome.
The e-CALLISTO FITS Analyzer adopts an intermediate approach: routine array
operations, RFI cleaning, background subtraction, masking, fitting, and export
are automated, while the researcher retains direct control over burst
selection, threshold adjustment, outlier rejection, and the choice of
fundamental or harmonic emission lane. This control is particularly useful for
Type~II bursts, whose emission may be broadened, fragmented, split into
multiple lanes, or partially obscured by interference.

The third contribution is the direct connection between visual inspection and
physical interpretation. Once a burst region is selected, the software extracts
a maximum-intensity backbone and fits its frequency--time evolution to estimate
the drift rate, which is then propagated into coronal shock-height and
shock-speed estimates under a selectable density model, each accompanied by a
quantified uncertainty. The tool therefore does not merely display e-CALLISTO
data, but provides a continuous and traceable path from raw or
background-subtracted dynamic spectra to physically meaningful quantities that
can be exported for publication, comparison, and statistical study.

Taken together, these features, cross-platform usability, open-source
implementation, interactive and uncertainty-quantified analysis, event-level
FITS merging, and built-in Type~II shock-parameter estimation make the
framework suitable for both individual case studies and future
catalogue-oriented work, while preserving transparency in the processing steps
used to obtain each measurement.

The remainder of this paper is organized as follows: Section 2 describes the properties of e-CALLISTO data. Section 3 details the software architecture and methodology. Section 4 demonstrates the analysis workflow using Type II bursts, and Section 5 concludes with a summary of the tool's capabilities.

\section{e-CALLISTO solar radio observations}
\label{sec:ecallisto_observations}

Solar radio dynamic spectra provide a direct view of transient energy release in the low and middle corona, where particle beams, shocks, and evolving magnetic structures generate coherent emission over metric and decametric wavelengths \citep{behlke2001solar}. The CALLISTO was developed as a compact, low-cost solar radio spectrometer that can be deployed widely, including at sites where access to large radio facilities is limited \citep{benz2005callisto}. The global deployment of many CALLISTO instruments forms the e-CALLISTO network \textcolor{red}{\footnote{https://www.e-callisto.org/}}, designed as a space-weather instrument array with the aim of achieving near-continuous monitoring of solar radio activity through longitudinal coverage \citet{Monstein_Csillaghy_Benz_2023}.

The significance of e-CALLISTO in space-weather field arises from its ability to capture radio signatures associated with eruptive phenomena and particle acceleration, including Type II, Type III, and Type IV bursts \citet{monstein2011catalog}. Because these emissions can occur over short time scales and may be visible only from certain geographic longitudes at a given time, a distributed network improves event capture probability and supports operational monitoring and retrospective studies. The e-CALLISTO project also emphasizes broad accessibility, combining scientific use with education, outreach, and long-term radio-frequency-interference monitoring at participating sites \citep{monstein2011catalog}.

\subsection{CALLISTO solar radio spectrometer}
\label{subsec:callisto_spectrometer}

CALLISTO is a programmable heterodyne receiver originally developed within the International Heliophysical Year and International Space Weather Initiative deployment framework, with an explicit focus on portability, low cost, and flexible observing programs \citep{davila2007international}. The instrument is designed to repeatedly sweep through a user-defined set of frequency channels and record intensity as a function of time and frequency, producing dynamic spectra suitable for burst detection and analysis \citep{benz2005callisto}.

In its standard configuration, CALLISTO operates in the metric range, with nominal coverage between approximately 45 and 870~MHz using a broadband tuner architecture \citep{benz2005callisto,benz2009world, Zucca2012Observations}. The observing program typically samples on the order of a few hundred channels per sweep, and the published network description reports a time resolution of about 0.25~s for spectra configured with 200 channels, together with a frequency resolution set by the receiver chain and channel spacing \citep{benz2009world, Zucca2012Observations}. \citet{benz2005callisto} note millisecond-scale integration and an overall dynamic range exceeding tens of decibels, which is sufficient for capturing both weak and intense burst features in routine monitoring.
A practical strength of CALLISTO is its configurability. Frequency channel lists can be defined to avoid persistent interference and target bands of interest, while optional frequency conversion hardware can extend the accessible range beyond the native tuner limits \citep{benz2005callisto}. This flexibility is central to network operation because it allows stations in different radio environments to tune their observing schedules and frequency plans while still producing standardized dynamic spectra products for the common archive \citep{Russu2015A}.

\subsection{Data from CALLISTO}
\label{subsec:callisto_data}

CALLISTO observations are stored as FITS-compatible ``FIT'' files and are commonly distributed in compressed form in the network archive. The network documentation summarizes the data product as a dynamic spectrum, where the horizontal axis is the time in UT, the vertical axis is the frequency in MHz, and the pixel values represent the instrument intensity in arbitrary digital units (ADU) \citep{adassuriya2023observation}. 

The CALLISTO operating manual describes each FITS file as comprising a primary header followed by the binary spectrogram data and two binary table components that encode the time axis and the frequency axis. This structure supports reliable reconstruction of the spectrogram in external analysis software because the axis information is carried along the intensity matrix, rather than being implied by filename conventions alone. 

In routine operation, data are written locally at the station and then uploaded to the central archive infrastructure. The e-CALLISTO site describes automated transfer of data from individual instruments to a central server and provides multiple access modes, including original FIT files, quicklook products, and derived daily overviews and light curves.

Archive naming conventions and quicklook products are designed to preserve essential provenance, including station identity and acquisition time. For example, the quicklook dataset documentation uses filenames of the form \texttt{STATION\_YYYYMMDD\_HHMMSS\_CODE.fit.gz}, which encodes the station name, observation date, start time, and an additional instrument/front-end descriptor. Such conventions simplify automated indexing and allow users to retrieve the corresponding raw FIT file for any plotted quicklook interval. 

\section{Software Implementation and Processing Workflow}
\label{sec:implementation_workflow}

\begin{figure*}
\centering
\includegraphics[width = 0.9\linewidth]{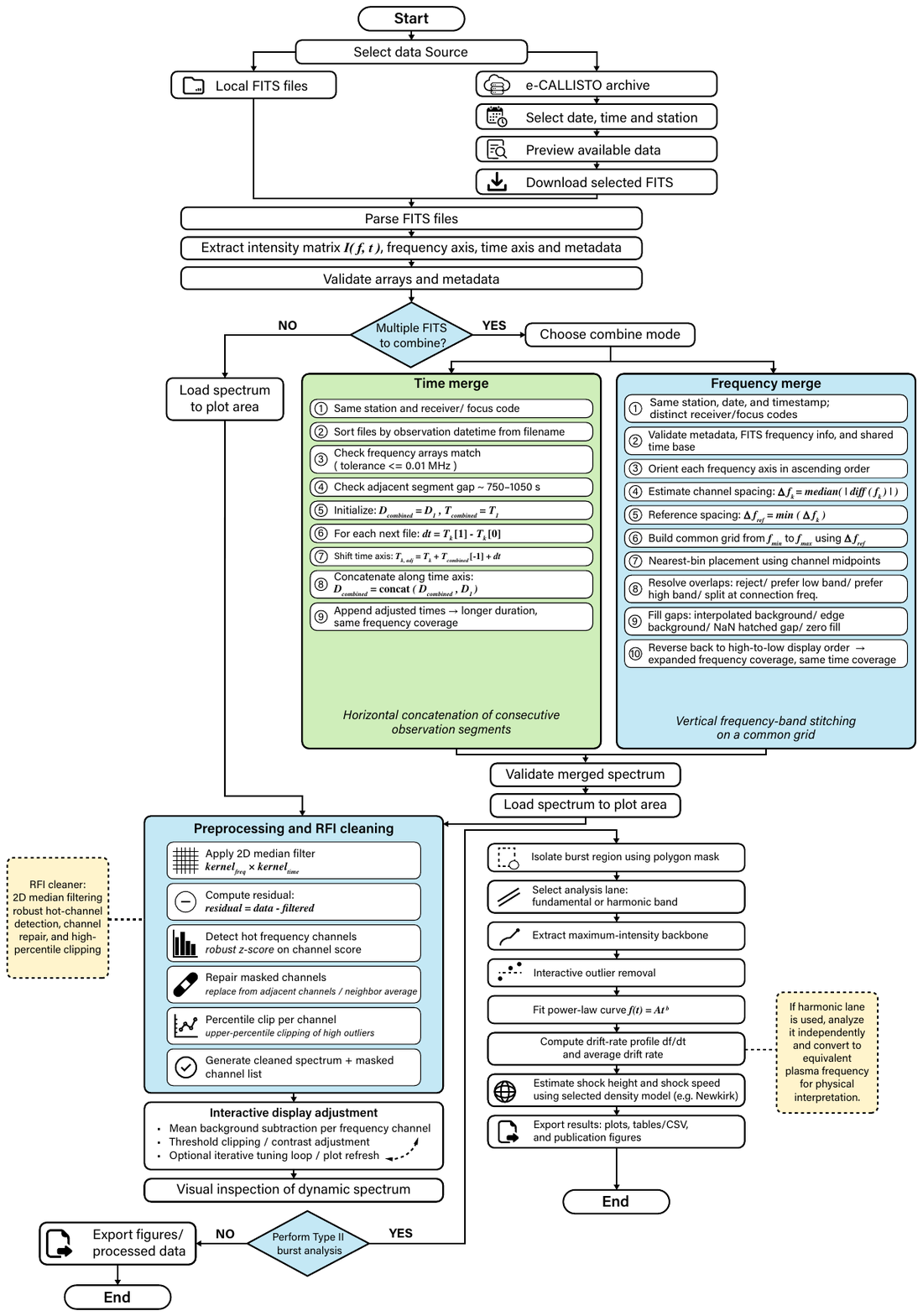}\
\caption{Overall workflow of the e-Callisto FITS Analyzer.}
\label{fig:workflow}
\end{figure*}

\subsection{Data model and FITS handling}
\label{subsec:data_model_fits}
The application adopts an in-memory representation that maps naturally to a solar radio dynamic spectrum. After import, each observation is modeled as a two-dimensional intensity matrix
\begin{equation}
I \in \mathbb{R}^{N_f \times N_t},
\end{equation}
where rows correspond to frequency channels and columns correspond to time samples. Two one-dimensional vectors define the coordinate system: the frequency vector $\{f_i\}_{i=1}^{N_f}$ (in MHz) and the time vector $\{t_j\}_{j=1}^{N_t}$ (in seconds). This representation is preserved across the entire workflow so that background removal, masking, and merging are expressed as well-defined operations on arrays with fixed axis semantics.

FITS files are parsed according to common e-CALLISTO conventions, in which the intensity matrix is stored in the primary data unit and the auxiliary coordinate information is provided through associated metadata and table content. The insertion pipeline supports both uncompressed FITS files and gzip-compressed FITS products, enabling the use of original station output as well as compressed archival products. When available, observation timing metadata are extracted from the FITS header to define an absolute reference time for later UT formatting (Figure \ref{fig:workflow}).

\subsection{Data acquisition and insertion}
\label{subsec:data_acquisition_insersion}
Data enter the processing pipeline through either local file selection or the integrated downloader. In both cases, the objective is to obtain a validated set of FITS inputs, extract the core arrays $(I,f,t)$, and populate the main plotting context while maintaining interactive responsiveness.

For remote acquisition, the downloader queries an archive directory for a user-selected date and filters the listing using station identifiers and time information encoded in filenames. Transfers are executed in a background worker so that the graphical interface remains responsive during network activity. Before import, an initial preview can be rendered to help confirm the station, time interval, and spectral coverage. Upon confirmation, the selected files are downloaded to a local workspace and passed through the same parsing path used for locally opened files, which ensures that subsequent processing and visualization are independent of the data source.

\subsection{FITS combination across time and frequency}

e-CALLISTO observations are commonly split across multiple FITS files, either because a long event extends over consecutive observing segments or because different receiver/focus settings cover different frequency sub-bands. To support event-level analysis, the application provides two combination modes: time merging and frequency merging. Both modes are applied only after the selected FITS files have been parsed, their intensity matrices and coordinate axes have been extracted, and their metadata have been checked. The combined output is a single dynamic spectrum with a consistent intensity matrix, frequency axis, and time axis for downstream visualization, cleaning, burst isolation, and Type~II burst analysis. The complete workflow is summarized in Figure~\ref{fig:workflow}.

In the time-merging mode, the selected files are treated as consecutive observing segments of the same frequency band. This mode is used when the files come from the same station and receiver/focus code, have matching frequency axes, and represent adjacent observation intervals. The files are first sorted according to the observation datetime encoded in the filename. The frequency axes are then compared with a tolerance of 0.01 MHz, and the separation between adjacent files is required to be consistent with one CALLISTO observing segment, here taken as approximately 750--1050 s. These checks prevent unrelated files or incompatible observing configurations from being merged into a single event.

Let $D_k$ be the dynamic spectrum of the $k$-th file, with frequency channels along rows and time samples along columns. Let $f_k$ and $T_k$ be the corresponding frequency and time arrays. The first file initializes the combined product as
\begin{equation}
D_{\rm comb}=D_1, \qquad
T_{\rm comb}=T_1, \qquad
f_{\rm comb}=f_1 .
\end{equation}
For each following file, the sampling interval is estimated from its time vector as
\begin{equation}
\Delta t_k = T_k[1] - T_k[0] .
\end{equation}
Because individual CALLISTO files commonly carry a time axis that begins near zero, the time vector of each following segment is shifted so that it starts immediately after the last sample of the current combined time series:
\begin{equation}
T_{k,{\rm adj}} = T_k + T_{\rm comb}[-1] + \Delta t_k .
\end{equation}
The intensity matrices are then concatenated along the time axis, while the adjusted time arrays are appended:
\begin{equation}
D_{\rm comb} \leftarrow \operatorname{concat}_{t}\!\left(D_{\rm comb},D_k\right),
\qquad
T_{\rm comb} \leftarrow \operatorname{concat}\!\left(T_{\rm comb},T_{k,{\rm adj}}\right).
\end{equation}
The result is a longer dynamic spectrum with unchanged frequency coverage but extended time duration. This is particularly useful for Type~II and Type~III bursts that continue across more than one 15-minute CALLISTO observing frame.

In the frequency-merging mode, the selected files are treated as simultaneous observations of different frequency sub-bands. This mode is used when the files have the same station, date, and timestamp, but distinct receiver/focus codes, and when their time arrays are compatible. The algorithm first validates the filename metadata, receiver/focus identifiers, FITS frequency information, and the shared time base. Each input frequency axis is then oriented in ascending order for grid construction.

For the $k$-th input band, the representative channel spacing is estimated from the median absolute difference between adjacent frequency channels:
\begin{equation}
\Delta f_k =
\operatorname{median}\left(\left|\operatorname{diff}(f_k)\right|\right).
\end{equation}
The common frequency grid is constructed using the smallest channel spacing among the selected files:
\begin{equation}
\Delta f_{\rm ref} = \min_k\left(\Delta f_k\right).
\end{equation}
If $f_{\min}$ and $f_{\max}$ are the minimum and maximum frequencies covered by all selected bands, the number of grid intervals is
\begin{equation}
N =
\operatorname{round}\left(
\frac{f_{\max}-f_{\min}}{\Delta f_{\rm ref}}
\right).
\end{equation}
The regular combined frequency grid is then defined as
\begin{equation}
f_{\rm grid} =
\operatorname{linspace}(f_{\min},f_{\max},N+1).
\end{equation}

Each source band is mapped into this common grid using nearest-bin placement based on channel midpoints. Therefore, valid source data are assigned to the nearest corresponding frequency rows of the combined grid rather than interpolated within the observed bands. This preserves the measured channel values while allowing bands with different channel spacings to be stitched into a single frequency axis.

When two input bands overlap, the result is controlled by the selected overlap policy. The overlap may be rejected, the lower-frequency band may be preferred, the higher-frequency band may be preferred, or the overlap may be split at a user-selected connection frequency. In split mode, rows below or equal to the connection frequency keep the previous band, while rows above it are replaced by the current band. If gaps remain between adjacent bands, they are filled according to the selected gap mode. The available gap treatments include interpolated background, average edge background, NaN hatched gap, or zero fill. In the background mode, valid rows on each side of the gap are used to estimate low-background traces, and the missing rows are filled by linear interpolation between the left and right background estimates.

After frequency-band stitching, the final combined array is returned to the application display convention, with the frequency axis ordered from high to low when required by the plotting interface. The output of frequency merging is therefore a dynamic spectrum with expanded frequency coverage and unchanged time coverage. Together, the time- and frequency-merging modes allow fragmented e-CALLISTO observations to be converted into a continuous event-level spectrum before cleaning, visualization, and quantitative burst analysis.

\subsection{RFI mitigation, background subtraction, and burst isolation}

Raw e-CALLISTO dynamic spectra commonly contain narrow-band radio-frequency interference, persistent channel offsets, and slowly varying background structure. The analyzer therefore applies a deterministic, statistics-based cleaning workflow before burst isolation and quantitative measurement. Let the imported dynamic spectrum be
\begin{equation}
I \in \mathbb{R}^{N_f \times N_t},
\end{equation}
where rows correspond to frequency channels and columns correspond to time samples. The cleaning procedure operates on this two-dimensional matrix while preserving the same frequency and time axes.

The first stage applies a two-dimensional median filter to suppress isolated bright pixels and small-scale impulsive artifacts. If $k_f$ and $k_t$ are the kernel sizes along the frequency and time axes, respectively, the filtered matrix is written as
\begin{equation}
I^{\rm med}_{i,j}
=
\operatorname{median}_{(u,v)\in \mathcal{K}_{f,t}}
I_{i+u,j+v},
\end{equation}
where $\mathcal{K}_{f,t}$ denotes the local median-filter window. In the implementation, even kernel sizes are increased to the nearest odd value, and nearest-neighbour boundary handling is used at the array edges. A residual matrix,
\begin{equation}
I^{\rm res}_{i,j}=I_{i,j}-I^{\rm med}_{i,j},
\end{equation}
is also computed as a diagnostic representation of local deviations from the median-filtered background.

The second stage identifies persistently contaminated frequency channels. For each frequency row $i$, the median channel level and the median absolute channel variation are calculated as
\begin{equation}
m_i = \operatorname{median}_{j}(I_{i,j}),
\end{equation}
and
\begin{equation}
a_i = \operatorname{median}_{j}\left|I_{i,j}-m_i\right|.
\end{equation}
A channel score is then defined as
\begin{equation}
q_i = |m_i| + a_i .
\end{equation}
This score is designed to flag channels that are either persistently bright or unusually variable. The score distribution is converted into a robust $z$-score using the median and median absolute deviation,
\begin{equation}
z_i =
0.6745\,\frac{q_i-\operatorname{median}(q)}
{\operatorname{median}\left|q-\operatorname{median}(q)\right|}.
\end{equation}
Channels satisfying
\begin{equation}
z_i > z_{\rm th}
\end{equation}
are marked as hot or contaminated channels, where the default threshold is $z_{\rm th}=6.0$. If the median absolute deviation is zero or not finite, the implementation falls back to a standard-deviation-based score.

The third stage repairs the masked channels using neighbouring frequency rows in the median-filtered matrix. For an interior masked channel $i$, the repaired row is
\begin{equation}
I^{\rm rep}_{i,j}
=
\frac{1}{2}
\left(
I^{\rm med}_{i-1,j}
+
I^{\rm med}_{i+1,j}
\right).
\end{equation}
For a masked channel at the boundary of the frequency axis, the nearest valid neighbouring row is copied. This step removes narrow-band contaminated channels while retaining the local time-dependent structure of neighbouring frequencies.

The fourth stage clips extreme high-intensity outliers on a row-by-row basis. For each frequency channel, the upper percentile value is computed as
\begin{equation}
P_i =
\operatorname{percentile}_{p}
\left(I^{\rm rep}_{i,:}\right),
\end{equation}
where the default value is $p=99.5$. The clipped output is then
\begin{equation}
I^{\rm rfi}_{i,j}
=
\min\left(I^{\rm rep}_{i,j},P_i\right).
\end{equation}
Only the high-intensity side is clipped, so that low-intensity structure and the general morphology of the burst are not artificially raised. The RFI-cleaning routine returns both the cleaned spectrum $I^{\rm rfi}$ and the list of masked frequency-channel indices, making the operation reproducible.

After RFI cleaning, the analyzer applies row-wise background subtraction. Invalid rows, including rows introduced as frequency gaps during frequency merging, are excluded from the baseline calculation and retained as undefined values. For each valid frequency channel, a baseline $b_i$ is estimated using one of three selectable methods:
\begin{equation}
b_i =
\begin{cases}
\operatorname{mean}_{j}(I^{\rm rfi}_{i,j}), & \text{mean mode},\\
\operatorname{median}_{j}(I^{\rm rfi}_{i,j}), & \text{median mode},\\
\operatorname{percentile}_{p_b}(I^{\rm rfi}_{i,:}), & \text{robust percentile mode}.
\end{cases}
\end{equation}
The default robust percentile is $p_b=25$, which estimates a low-background level while reducing the influence of bright burst emission. The background-subtracted spectrum is
\begin{equation}
\widetilde{I}_{i,j}=I^{\rm rfi}_{i,j}-b_i .
\end{equation}

An optional noise-equalization step can be applied after background subtraction. For each valid frequency row, a robust noise scale is estimated from the median absolute deviation:
\begin{equation}
\sigma_i =
1.4826\,\operatorname{median}_{j}
\left|
\widetilde{I}_{i,j}
-
\operatorname{median}_{j}(\widetilde{I}_{i,j})
\right|.
\end{equation}
If this estimate is not finite or is too small, the implementation falls back first to the interquartile range divided by 1.349, then to the standard deviation, and finally to unity. A target scale $\sigma_{\rm tar}$ is selected from a low percentile of the valid row-scale distribution, with a default percentile of 25. The row-wise equalization factor is
\begin{equation}
\eta_i =
\frac{\sigma_{\rm tar}}{\sigma_i}.
\end{equation}
When the attenuate-only option is enabled, $\eta_i$ is limited to a maximum value of one:
\begin{equation}
\eta_i \leftarrow \min(\eta_i,1).
\end{equation}
The equalized spectrum is then
\begin{equation}
I^{\rm eq}_{i,j}=\eta_i\widetilde{I}_{i,j}.
\end{equation}
This option suppresses unusually noisy rows without amplifying quiet rows, making it useful when residual channel-dependent noise remains after RFI cleaning.

For visualization and interactive inspection, the user may then apply lower and upper display thresholds. If $L$ and $H$ are the selected clipping limits, the displayed or thresholded matrix is
\begin{equation}
I^{*}_{i,j}
=
\min\left[
\max\left(I^{\rm eq}_{i,j},L\right),H
\right],
\end{equation}
or equivalently $I^{*}_{i,j}=\min[\max(\widetilde{I}_{i,j},L),H]$ when noise equalization is not applied. These thresholds are adjustable in the graphical interface, allowing the user to tune the contrast iteratively without reloading the data.

Burst isolation is implemented as an interactive masking operation in the time-frequency plane. A user-defined closed region is traced around the feature of interest, producing a polygon $P$ in plot coordinates. This region is mapped onto the discrete grid to form a Boolean mask,
\begin{equation}
M_{i,j} =
\begin{cases}
1, & (t_j,f_i)\in P,\\
0, & \text{otherwise}.
\end{cases}
\end{equation}
The isolated burst product is then computed by element-wise masking:
\begin{equation}
I_{\rm iso}=M\odot I^{*},
\end{equation}
where $\odot$ denotes the Hadamard product. Pixels outside the selected region are suppressed, yielding a burst-only matrix that can be used for subsequent backbone extraction, drift-rate fitting, and shock-parameter estimation.

The present cleaning method is intended as RFI mitigation and background suppression in instrumental intensity units. It does not by itself provide absolute radiometric calibration into solar flux units. Full calibration into SFU requires station-specific calibration information and an appropriate sky or system-temperature model, which is identified as a future extension of the software.

\subsubsection{Relative digit-to-dB scaling and calibration status}

The intensity values stored in the e-CALLISTO FITS files are treated in the analyzer as native instrumental values in ADU, also referred to as digits. For visualization and relative intensity comparison, the main display converts these digit values to a CALLISTO-style decibel scale using a linear digit-to-dB mapping rather than a logarithmic power-ratio conversion. In the main viewer, the conversion is written as
\begin{equation}
I_{\rm dB}
=
\left(I_{\rm ADU}-D_{\rm cold}\right)\,\frac{2500}{256 \times 25.4},
\end{equation}
where $D_{\rm cold}$ is taken from the current lower noise or clipping threshold. This gives a scale factor of approximately $0.3846~{\rm dB}$ per ADU. The conversion is applied after optional row-wise background subtraction and display clipping, and therefore represents a relative intensity scale referenced to the selected background level.

In the batch \texttt{median\_dB} mode, the reference level is taken from the median of each frequency row rather than from the display threshold. The corresponding conversion is
\begin{equation}
I_{\rm dB}
=
\left(I_{\rm ADU}-\operatorname{median}_{j}(I_{{\rm ADU},i,j})\right)\,\frac{2500}{255 \times 25.4},
\end{equation}
which gives approximately $0.3861~{\rm dB}$ per ADU. For full-day spectral overviews, the same scaling is applied after subtracting a day-wide median baseline for each frequency channel. These options provide a consistent relative background-subtracted dB representation for visual inspection, burst isolation, and comparison of intensity structure within the same observing configuration.

This procedure should not be interpreted as absolute radiometric calibration. The present version does not convert the spectra into solar flux units (SFU) or antenna temperature. Absolute calibration would require additional station-specific information, including antenna gain as a function of frequency and pointing, receiver response, system temperature, and an appropriate sky or calibration-source model. Since these quantities are not uniformly available for all e-CALLISTO stations, the current release reports cleaned and background-subtracted spectra in instrumental units or relative dB units. Full SFU calibration using a sky model is therefore treated as a future extension rather than a capability of the present version.

\subsection{Data visualization}
\label{subsec:data_visualization}
Visualization is built on an embedded plotting canvas that renders the dynamic spectrum as a two-dimensional image with physically meaningful axes. The intensity matrix is displayed using an extent that maps matrix indices onto the time and frequency vectors, so cursor readout and interactive selections operate in seconds and MHz rather than pixel coordinates.

Time labeling can be expressed either as elapsed seconds or as UT. When UT mode is enabled, an observation start time $t_0$ extracted from FITS metadata is used to format displayed time as
\begin{equation}
t_{\mathrm{UT}} = t_0 + t.
\end{equation}
Color scaling is managed through a dedicated colorbar axis and a selectable colormap, enabling consistent inspection across raw, background-subtracted, and isolated-burst views. The visualization layer separates data state from display state so that plot updates can be performed without discarding the user’s current view limits.

\subsection{User interaction and control layer}
\label{subsec:user_interaction_control}
The user interaction layer is designed to support rapid, iterative exploration. Background suppression thresholds are exposed through continuous controls and applied directly to the in-memory representation, enabling immediate feedback during tuning. Standard zooming and panning interactions allow users to focus on specific time--frequency regions, and view limits can be preserved across plot refreshes so that users maintain spatial context while adjusting thresholds, colormaps, or axis formatting.

Interactive isolation tools are integrated into the same plotting context so that selection and visualization remain consistent. The cursor readout provides time, frequency, and intensity values at the pointer location, supporting manual inspection of burst lanes. To support iterative analysis, the application maintains an undo and redo history that captures both data-state changes and view state, which reduces friction when exploring multiple parameter choices.

\subsection{Extensibility and modular design}
\label{subsec:extensibility_modularity}
The software is structured as modular components that share a common graphical framework while remaining loosely coupled at the data level. The core radio pipeline focuses on FITS insertion, cleaning, visualization, and burst isolation. Space-weather context products are integrated as separate modules launched from the main interface. GOES X-ray flux visualization is implemented as a dedicated component that retrieves and plots time-series data, while SOHO/LASCO CME browsing is provided through an independent viewer tailored to catalog access and event inspection.

This modular organization supports extension in two directions. First, the radio analysis workflow can remain stable while new context modules are added. Second, each external data stream can evolve its own parsing, caching, and plotting logic without forcing changes to FITS handling or the radio processing pipeline.

\subsection{Software dependencies and environment}
\label{subsec:dependencies_environment}
The application is implemented in Python and relies on established scientific and graphical libraries to enable cross-platform execution, FITS I/O, interactive visualization, numerical processing, and access to external space-weather resources. Dependencies were selected to balance stability, performance, and maintainability, while keeping installation practical for end users (Table \ref{tab:deps}).

\begin{table*}
\centering
\caption{Principal software dependencies and their roles. Versions should be reported from the reference environment used in the experiments.}
\label{tab:deps}
\centering
\begin{tabular}
{|p{0.15\textwidth}|p{0.50\textwidth}|p{0.07\textwidth}|}
\hline
\multicolumn{1}{|c|}{\textbf{Dependency}} &
\multicolumn{1}{c|}{\textbf{Role in the software}} &
\multicolumn{1}{c|}{\textbf{Version used}} \\
\hline
\hline
Python & Runtime environment & 3.13.4 \\
PySide6 & Desktop GUI framework and Qt bindings & 6.10.1 \\
NumPy & Array model and numerical operations on spectra & 2.3.5 \\
Matplotlib & Embedded plotting, colormaps, interaction, and figure export & 3.10.7 \\
PyQtGraph & High-performance plotting & 0.14.0 \\
Astropy & FITS input/output, header handling, and astronomy data utilities & 7.2.0 \\
SciPy & Numerical utilities used in analysis routines & 1.16.3 \\
OpenPyXL & Spreadsheet export & 3.1.5 \\
Requests & HTTP retrieval for remote data sources and update checks & 2.32.5 \\
BeautifulSoup4 & HTML parsing for server listings and catalogs & 4.14.3 \\
netCDF4 & GOES data insertion & 1.7.3 \\
cftime & Time handling for netCDF and GOES products & 1.6.5 \\
SunPy & Multi-mission solar archive query, download, plotting, and analysis & 7.1.0 \\
scikit-learn & Optional analysis utilities where applicable & 1.7.2 \\
setuptools & Build and distribution support & 80.9.0 \\
\hline
\end{tabular}

\end{table*}

For reproducibility, analysis can be accompanied by a record of the computational environment used to generate exported figures and derived data products. This is typically achieved by archiving a dependency manifest or an environment specification alongside the processed dataset and manuscript materials.

Apart from that, the e-CALLISTO FITS Analyzer natively works on Windows, macOS and Linux (Debian/Ubuntu) and native installers are available to download and use via the official e-CALLISTO website\footnote{https://www.e-callisto.org/Software/Callisto-Software.html}. Also, the source code is published at GitHub\footnote{https://github.com/SaanDev/e-Callisto\_FITS\_Analyzer}.

\subsection{Design considerations and limitations}
\label{subsec:design_considerations_limitations}
Several design choices prioritize interactive research use while maintaining clear separation between data transformations and visualization. The application maintains a single canonical in-memory representation of the spectrum, and plot refreshes are derived from this state so that processing steps remain traceable. Network activity is delegated to worker threads to avoid blocking the interface, and plotting updates are managed through the GUI event loop to reduce instability during interactive sessions.

The methodology also imposes practical limitations. Combination across time and frequency depends on compatibility assumptions that generally hold for standard e-CALLISTO products but may fail for atypical station configurations or inconsistent acquisition settings. Time merging constructs a continuous axis through cumulative offsets and does not, by itself, guarantee correction for gaps, overlaps, or station clock irregularities. Frequency merging assumes compatible time sampling across selected files, and reliability depends on consistency in metadata and file naming conventions.

Background subtraction is intentionally model-light. Mean background subtraction and threshold clipping perform well for many burst morphologies but can suppress weak diffuse emission and may alter the apparent intensity distribution when the background varies rapidly. Burst isolation depends on a user-defined region and on the mapping between continuous plot coordinates and the discrete data grid, which can introduce selection sensitivity near burst boundaries. Finally, the present release provides instrumental-unit and relative digit-to-dB representations, but does not perform absolute radiometric calibration into SFU or antenna temperature. Such calibration requires station-specific antenna gain, receiver response, system-temperature information, and a sky or calibration-source model. Incorporating these calibration-aware products is planned as a future extension.

\section{analysis of Type II bursts}
\label{sec:typeII_analysis}

Type II solar radio bursts provide one of the most direct radio diagnostics of coronal shock propagation \citep{mann1995characteristics}. In dynamic spectra, they appear as slowly drifting lanes that trace the plasma frequency (or its harmonic) as the shock traverses decreasing coronal density \citep{tsap2020evolution}. Quantifying the lane trajectory therefore enables physically meaningful parameters to be inferred, including the frequency drift rate, shock formation height, and shock speed under an assumed coronal density model \citep{late1965nature, gopalswamy2013height}. In metric observations, Type II bursts often show both a fundamental band and a harmonic band. The fundamental emission is expected near the local plasma frequency \(f_{\mathrm{pe}}\), while the harmonic emission occurs near \(2f_{\mathrm{pe}}\) \citep{cane2005solar}. In the present workflow, calculations are performed using the fundamental band whenever it is sufficiently clear. When the fundamental lane is weak, fragmented, or obscured, the harmonic lane can be selected and converted to the equivalent fundamental frequency by \(f_{\mathrm{pe}} \approx f_{\mathrm{H}}/2\) prior to physical parameter estimation. This selection is performed directly in the analyzer interface, allowing the measurement to remain consistent with the visible lane morphology. \par

\begin{figure*}
\centering
\includegraphics[width = \linewidth]{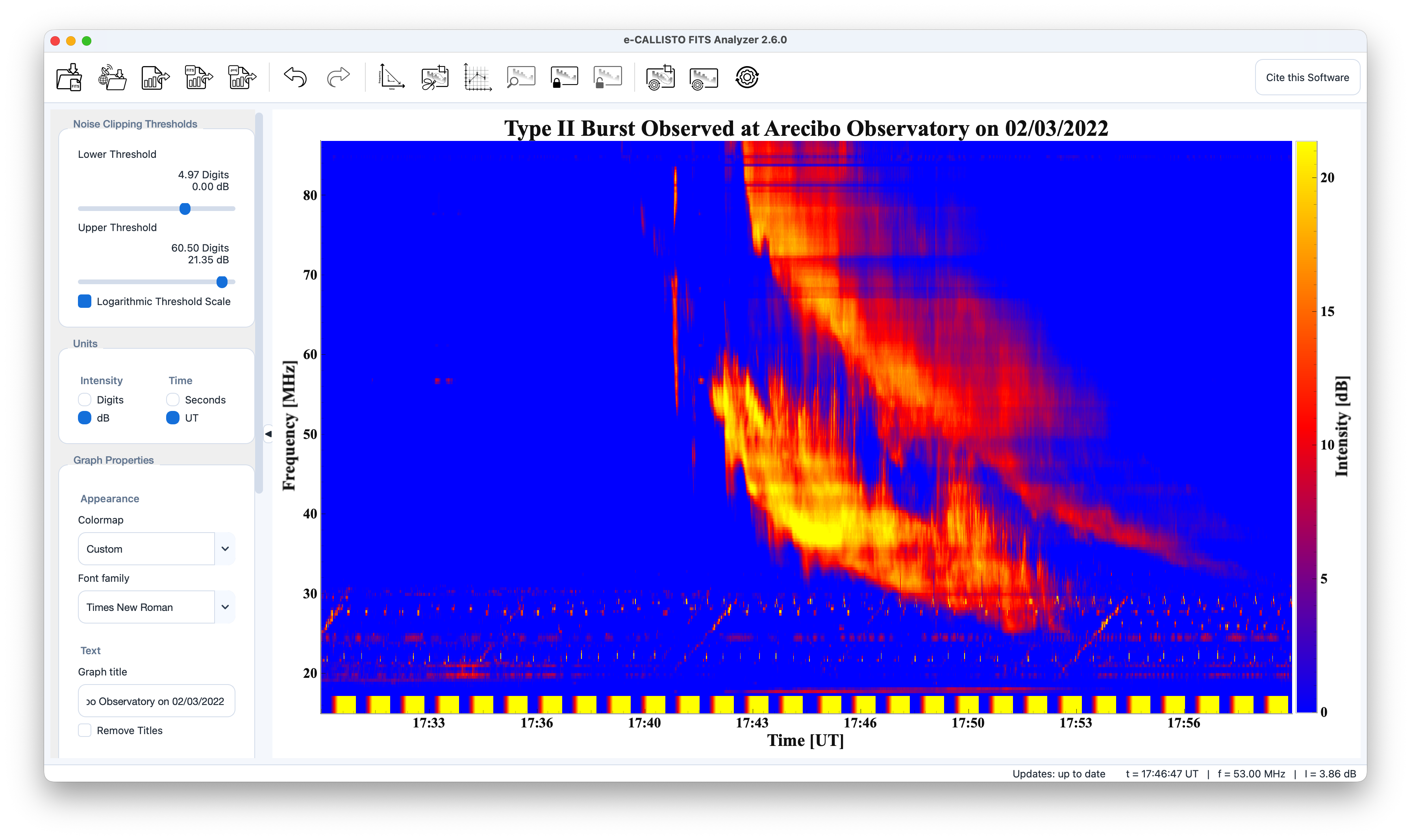}
\caption{The main application window showing a Type II solar radio burst observed at Arecibo Observatory on 2 March 2022 after background subtraction.}
\label{fig:mainwindow}
\end{figure*}

\subsection{Maximum-intensity backbone after isolation}
\label{subsec:max_intensity}
After performing background subtraction and isolating the burst, the next step in the analysis is to obtain a one-dimensional representation of the burst that reflects how its characteristic frequency changes over time. The application creates this representation by extracting a backbone curve from the isolated intensity matrix. Let \(I \in \mathbb{R}^{N_f \times N_t}\) be the processed spectrum and \(M \in \{0,1\}^{N_f \times N_t}\) be the isolation mask, where \(M_{i,j}=1\) for pixels that remain within the chosen burst region. For each time index \(j\), the backbone is defined by selecting the frequency index that maximizes the intensity after masking:
\begin{equation}
i^\star(j) = \arg\max_{i}\,\big(M_{i,j}\, I_{i,j}\big),
\qquad
f_{\max}(t_j) = f_{i^\star(j)}.
\end{equation}
This procedure converts the two-dimensional burst structure into a single curve \(f_{\max}(t)\) that follows the most intense ridge of emission at each time step \citep{liyanage2025determination}. The maximum-intensity backbone is well suited to Type II analysis because the lane is often band-limited but not infinitesimally thin, and the local intensity maximum provides a practical proxy for the lane center even when the emission is broadened by instrumental response, scattering, or fine structure. \par

Residual artifacts may still appear as outliers in the extracted backbone. These outliers commonly arise from incomplete suppression of radio-frequency interference, leakage from the harmonic lane, or isolated bright pixels remaining after clipping. Instead of relying solely on automatic heuristics, the tool provides an interactive outlier removal step in which the user selects spurious points directly on the backbone plot and removes them immediately. This interaction is particularly effective because outliers are typically separated geometrically from the smooth lane trend in the \((t,f)\) plane, and manual selection preserves scientifically meaningful points that may be rejected by overly aggressive automated filters. The cleaned backbone then forms the input to parametric fitting and drift-rate estimation. \par

\subsection{Frequency drift-rate calculation}
\label{subsec:drift_rate}
The frequency drift rate is a key observable of Type II bursts because it encodes the rate at which the shock-driven emission region moves through the coronal density gradient. After outlier removal, the backbone points are fitted using a power-law model,
\begin{equation}
\label{eqn:power_law}
f(t) = A\, t^{b},
\end{equation}
where \(A\) and \(b\) are fitting parameters and \(t\) denotes the elapsed time along the selected burst segment \citep{liyanage2025determination}. The instantaneous drift rate is obtained from the derivative of the fitted curve,
\begin{equation}
\label{eqn:inst_drift}
\frac{df}{dt} = A\,b\, t^{(b-1)}.
\end{equation}
Since Type II lanes may exhibit short-scale irregularities even after cleaning, the application reports an average drift rate that summarizes the overall lane trend across the selected interval. If the fitted curve is evaluated at \(N\) time samples \(\{t_j\}\), the mean absolute drift rate can be expressed as
\begin{equation}
\label{eqn:mean_drift}
\overline{D} = \frac{1}{N}\sum_{j=1}^{N}\left|\left.\frac{df}{dt}\right|_{t=t_j}\right|.
\end{equation}
This approach reduces sensitivity to local measurement noise while preserving the physically relevant monotonic drift of the Type II burst \citep{liyanage2025determination}. \par

\begin{figure*}
    \centering
    \begin{subfigure}[t]{0.48\textwidth}
        \centering
        \includegraphics[width=\linewidth]{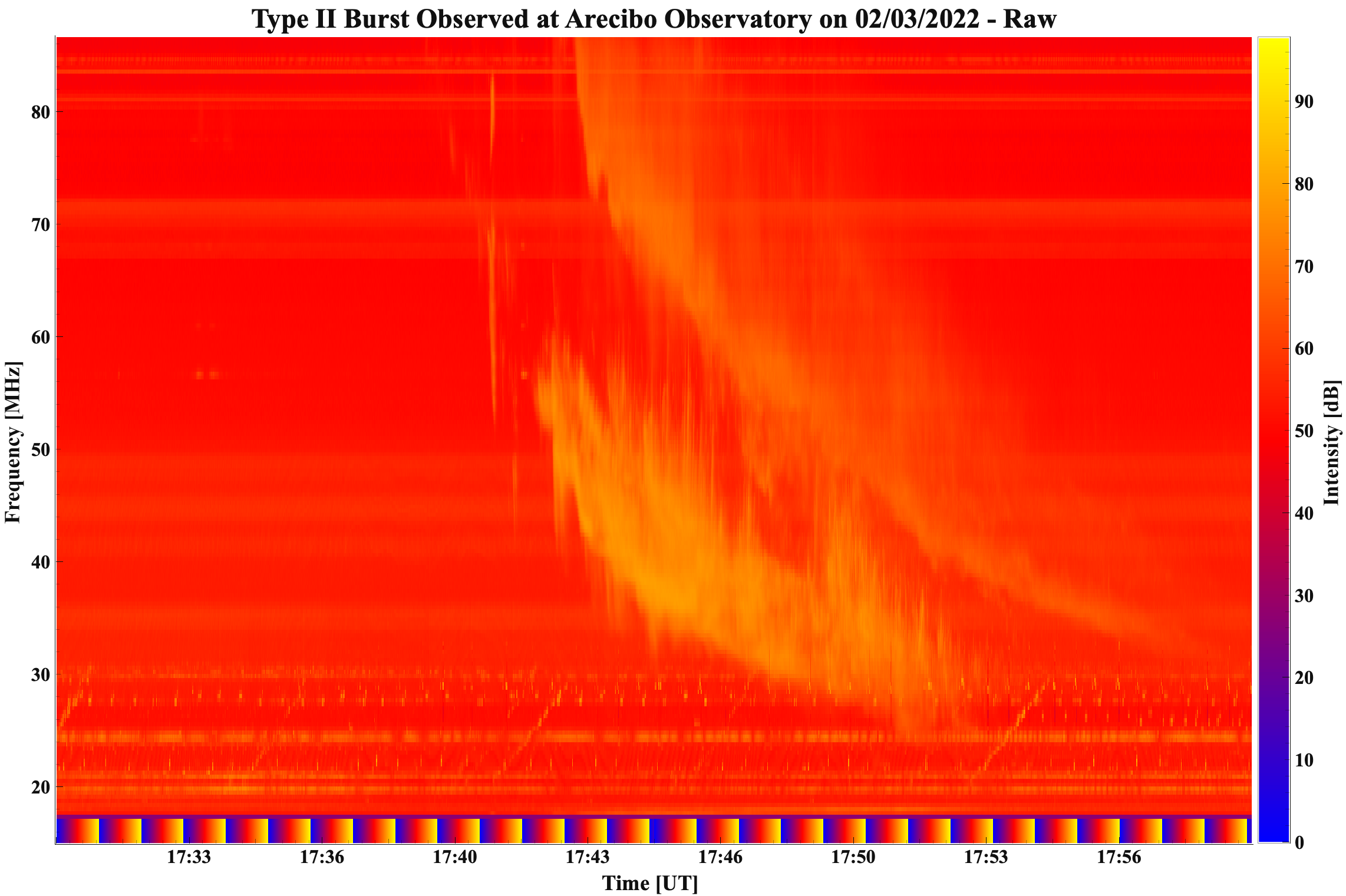}
        \caption{Raw dynamic spectrum of the Type II solar radio burst recorded by the CALLISTO instrument at Arecibo Observatory on 02/03/2022.}
        \label{fig:raw}
    \end{subfigure}
    \hfill
    \begin{subfigure}[t]{0.48\textwidth}
        \centering
        \includegraphics[width=\linewidth]{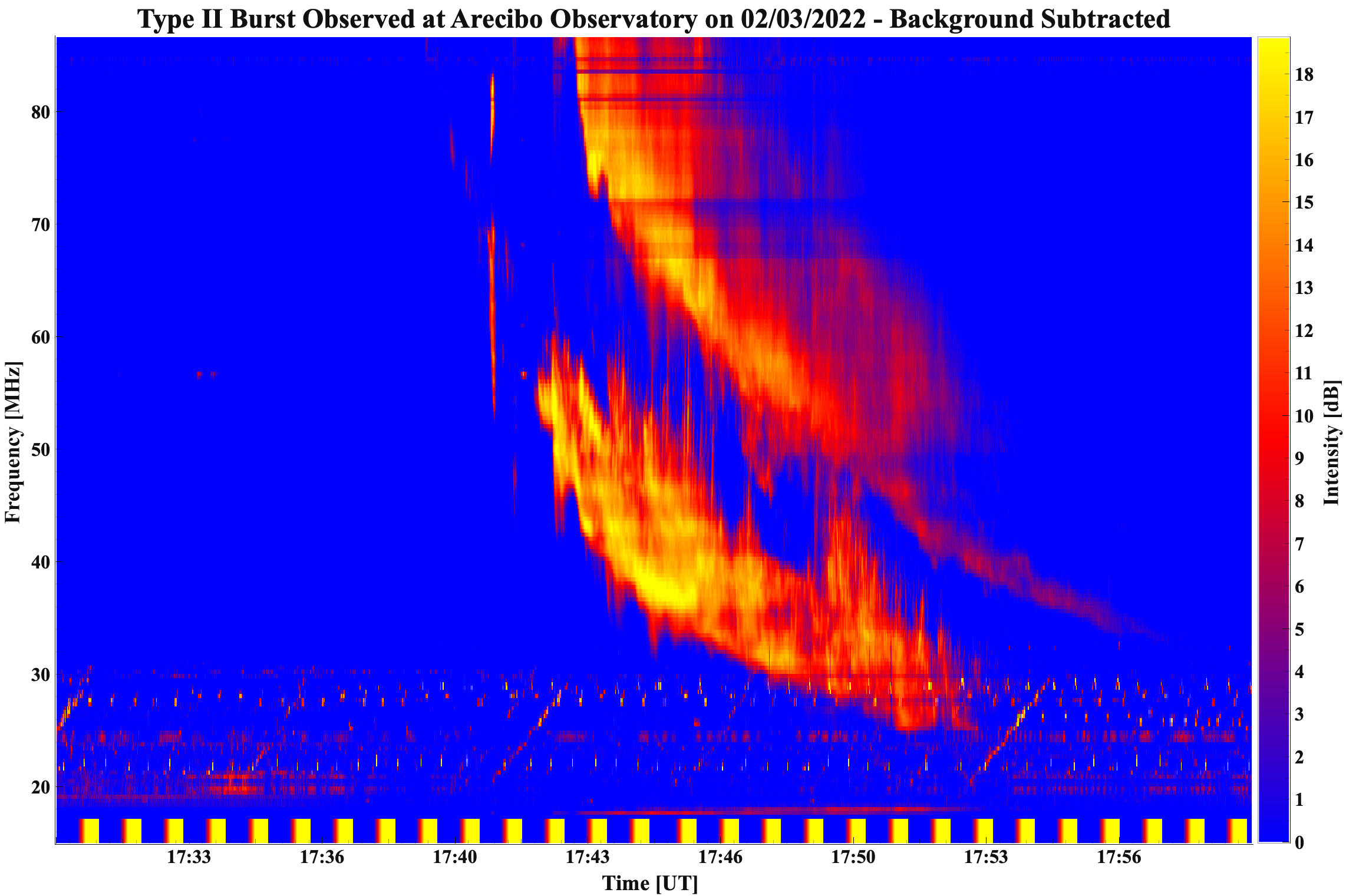}
        \caption{Background subtracted dynamic spectrum of the Type II solar radio burst recorded by the CALLISTO instrument at Arecibo Observatory on 02/03/2022.}
        \label{fig:bgs}
    \end{subfigure}

    \begin{subfigure}[t]{0.48\textwidth}
        \centering
        \includegraphics[width=\linewidth]{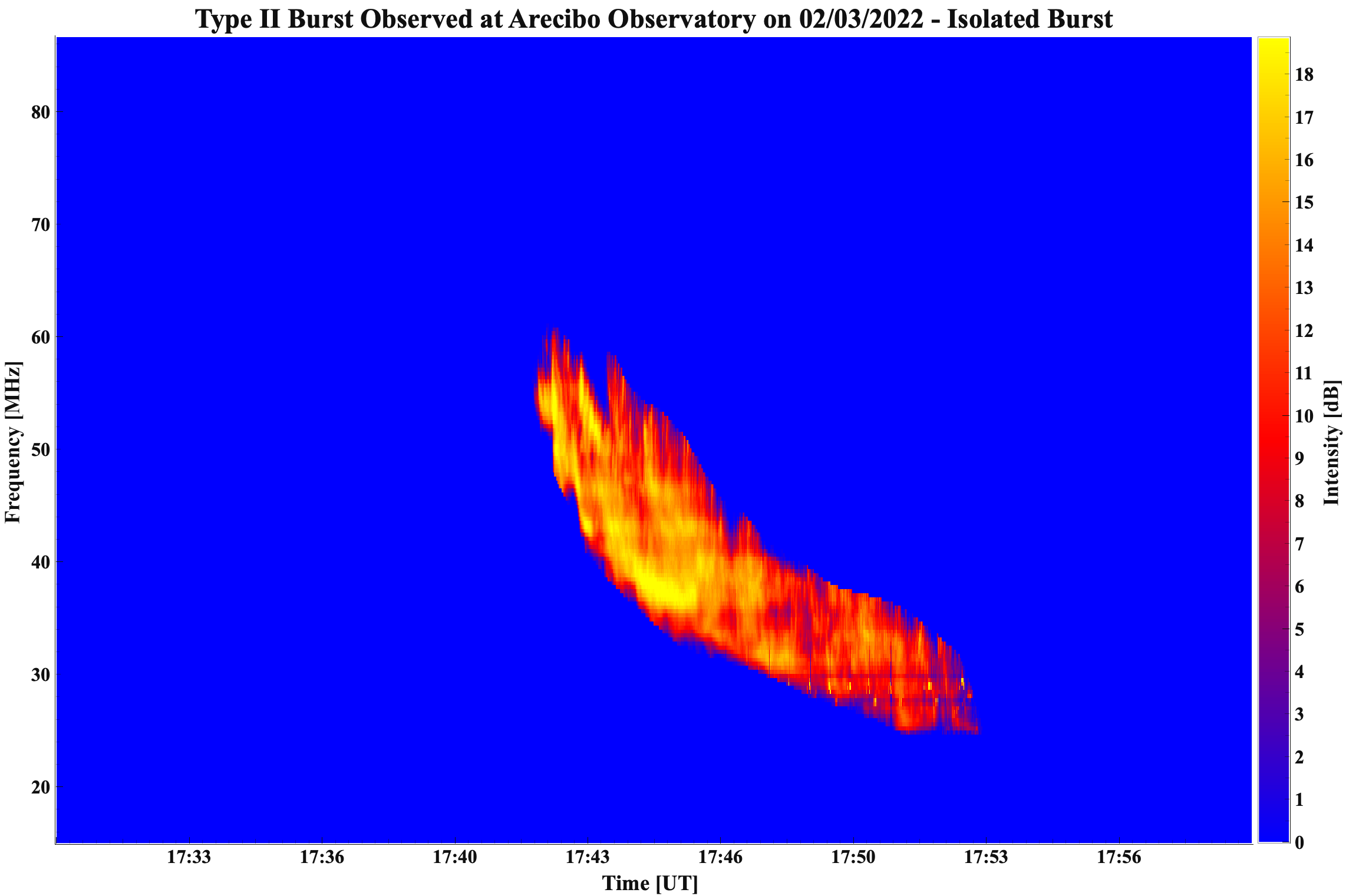}
        \caption{Isolated fundamental band of  dynamic spectrum of the Type II solar radio burst recorded by the CALLISTO instrument at Arecibo Observatory on 02/03/2022.}
        \label{fig:IB}
    \end{subfigure}
    \hfill
    \begin{subfigure}[t]{0.48\textwidth}
        \centering
        \includegraphics[width=\linewidth]{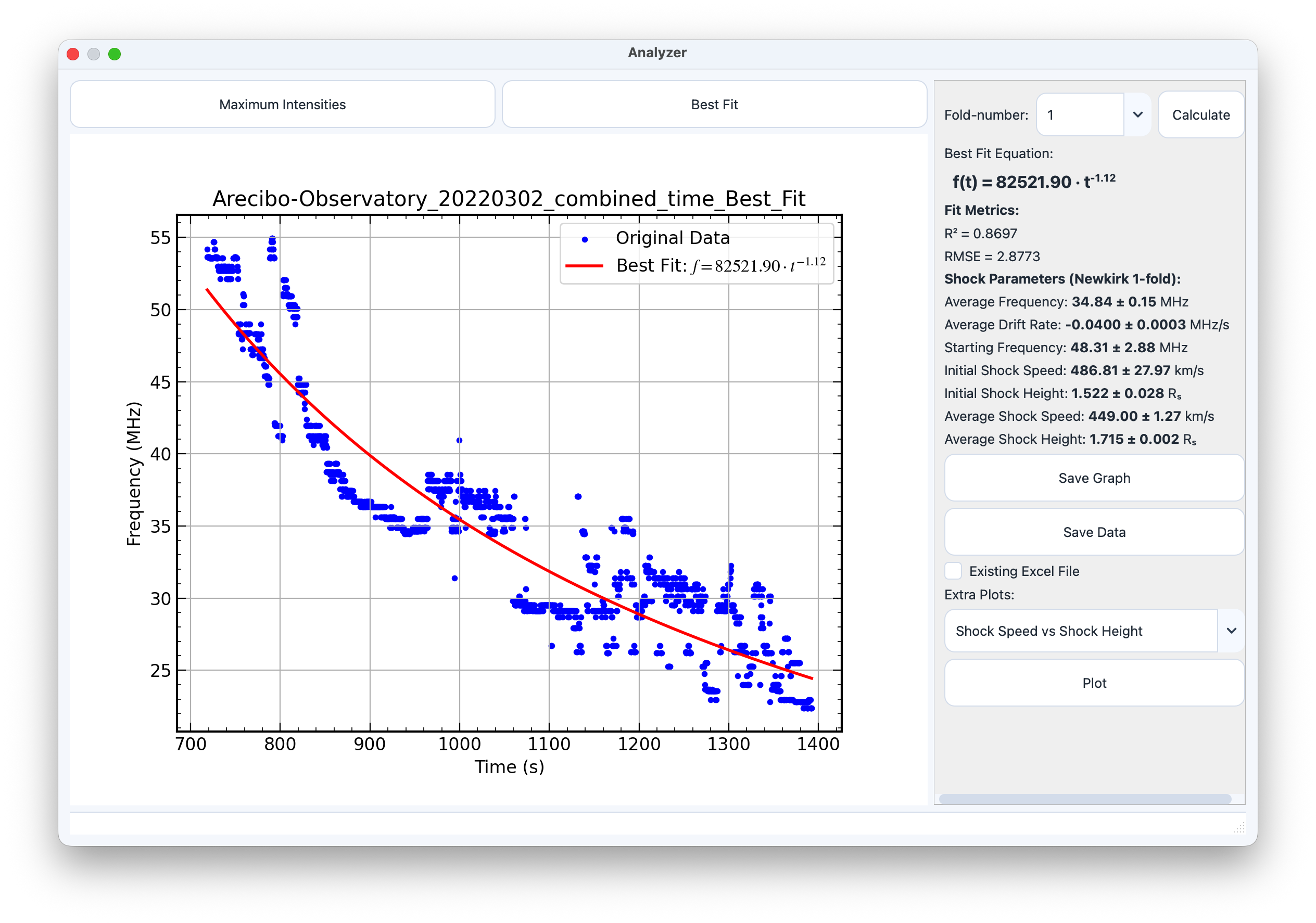}
        \caption{Analyzer window showing the maximum-intensity backbone and the derived shock parameters for the same burst, calculated using the one-fold Newkirk electron density model.}
        \label{fig:analyzer}
    \end{subfigure}

    \caption{Processing stages of the e-CALLISTO FITS Analyzer for a Type II solar radio burst observed by CALLISTO at Arecibo Observatory on 02/03/2022: (a) raw dynamic spectrum, (b) background-subtracted dynamic spectrum, (c) isolated fundamental band, and (d) analyzer window showing the maximum-intensity backbone and derived shock parameters computed using the one-fold Newkirk electron density model.}
    \label{fig:workflow_images}
\end{figure*}

\subsection{Shock-parameter estimation}
\label{subsec:shock_parameters}
To determine the coronal shock parameters the measured drift rate and electron density model \(n_e(R)\) have been used. Under the one-fold Newkirk density model, the coronal electron density varies with heliocentric distance \(R\) as
\begin{equation}
n_e(R) = n_0\,10^{\alpha\,(R_\odot/R)},
\end{equation}
where \(n_0=4.2\times10^{4} \ \mathrm{cm^{-3}}\) is the electron density near to the solar surface, \(\alpha=4.32\) is a fitting constant, and \(R_\odot=696340 \ \mathrm{km}\) is the solar radius \citep{1961ApJ...133..983N}. The expression generalises to the $n$-fold Newkirk model by scaling the surface
density with an integer factor $n$, $n_e(R)=n\,n_0\,10^{\alpha(R_\odot/R)}$, which
amounts to replacing $n_0$ with $n\,n_0$ in Equations~\ref{eqn:shock_height}
and~\ref{eqn:shock_speed}; the results reported in this work use the one-fold model
($n=1$). 
\\
The plasma frequency can be written in MHz as
\begin{equation}
f_{\mathrm{pe}} = \kappa\sqrt{n_e},
\end{equation}
where \(\kappa\approx 8.978 \times 10^{-3} \ \mathrm{MHz\ cm^{\frac{3}{2}}}\) is a constant that collects physical constants and unit conversions \citep{nicholson1983introduction}. Combining the density model with the plasma-frequency relation yields an explicit expression for the shock source height \(R_s\) as a function of plasma frequency \citep{liyanage2025determination}:
\begin{equation}
\label{eqn:shock_height}
R_s =\frac{\alpha R_\odot \ln(10)}
{\ln\!\left(\dfrac{f_{\mathrm{pe}}^{2}}{n_0 \kappa^{2}}\right)}
\end{equation}
Differentiating this expression with respect to time gives the shock speed \(V_s = dR_s/dt\), which can be written in terms of the frequency drift rate \citep{liyanage2025determination}:
\begin{equation}
\label{eqn:shock_speed}
V_s =
\frac{2\alpha R_\odot \ln(10)}
{\Lambda^2f_{\mathrm{pe}}\,}
\left|\frac{df_{\mathrm{pe}}}{dt}\right|,
\qquad
  \Lambda=\ln\!\Big(\frac{f_{\rm pe}^{2}}{n_{0}\kappa^{2}}\Big),
\end{equation}

In the application, $f_{\rm pe}$ is taken from the selected band. For a
fundamental band the observed band frequency is treated directly as
$f_{\rm pe}$. For a harmonic band the lane is fitted in the same way and
mapped onto the plasma-frequency scale through the standard harmonic
relation $f_{\rm H}\approx 2f_{\rm pe}$ \citep{cane2005solar}, so that
$f_{\rm pe}=f_{\rm H}/2$ and, by differentiation,
$\mathrm{d}f_{\rm pe}/\mathrm{d}t=\tfrac{1}{2}\,\mathrm{d}f_{\rm H}/\mathrm{d}t$.
The harmonic-derived plasma-frequency drift is therefore obtained
independently from the harmonic backbone, not copied from the fundamental;
it coincides with the fundamental-derived value only to the extent that the
harmonic relation holds. The drift rate is obtained from the fitted
derivative, while the source height is estimated from the starting frequency
(for example, a starting-frequency estimate derived from the upper portion
of the band). These steps produce a self-consistent set of shock parameters
tied directly to the user-verified band morphology. As with the drift rate (Equation~\ref{eqn:mean_drift}), the shock height and speed are reported both at the burst onset and as averages over the fitted lane. The onset values are evaluated at the starting frequency $f_s$ (the upper-percentile estimate described above), whereas the average values are the means of the per-sample quantities along the fitted track,
\begin{equation}
\bar{R}_s=\frac{1}{N}\sum_{j=1}^{N} R_s(f_j),\qquad
\bar{V}_s=\frac{1}{N}\sum_{j=1}^{N} V_s(f_j),
\label{eq:meanshock}
\end{equation}
evaluated over the $N$ samples spanning the analysed interval.

The robustness of the harmonic branch follows directly from the
shock-speed expression. Recasting Equation~\ref{eqn:shock_speed} in terms of the logarithmic
drift,
\begin{equation}
  V_{s}=\frac{2\,\alpha\,R_{\odot}\ln 10}{\Lambda^{2}}
        \left|\frac{\mathrm{d}\ln f_{\rm pe}}{\mathrm{d}t}\right|,
  \qquad
  \Lambda=\ln\!\Big(\frac{f_{\rm pe}^{2}}{n_{0}\kappa^{2}}\Big),
  \label{eq:vs_logdrift}
\end{equation}
makes explicit that the burst enters the shock speed only through
$\mathrm{d}\ln f_{\rm pe}/\mathrm{d}t$. This logarithmic drift is invariant
under a constant rescaling of the frequency axis: if the true harmonic
ratio is $r$ rather than exactly $2$, both $f_{\rm pe}$ and
$\mathrm{d}f_{\rm pe}/\mathrm{d}t$ are scaled by the same factor $r/2$ and
$\mathrm{d}\ln f_{\rm pe}/\mathrm{d}t$ is unchanged. The harmonic conversion
thus leaves the drift-derived part of the shock speed unaffected; the only
residual sensitivity is to the absolute frequency scale, entering weakly
through $\Lambda$. For the present event, a representative $\pm5\%$
departure of the harmonic ratio from $2$ ($r=1.9$--$2.1$) changes the
inferred shock speed by $\lesssim 3.5\%$ and the shock height by
$\lesssim 1.7\%$, comparable to the statistical uncertainties quoted in
Section~\ref{sec:uncertainty}. Larger, time-dependent departures can arise
from band splitting, shock geometry, emission-height differences, and
propagation effects \citep{cane2005solar,tsap2020evolution,vrsnak2008}, which a single
global ratio does not capture. For this reason the fundamental band is used
whenever it is sufficiently clear, the harmonic branch serving as a fallback
when the fundamental lane is weak, fragmented, or obscured; in that regime
the harmonic is often the cleaner tracer, as emission near $2f_{\rm pe}$
propagates more freely than fundamental emission near the local plasma
level. The Type~II parameters reported in Section~\ref{subsec:example_usage} were
derived from the fundamental band (Figure \ref{fig:analyzer}) and are therefore independent of the harmonic-ratio assumption.

\subsection{Uncertainty estimation}
\label{sec:uncertainty}

Every reported quantity is accompanied by an uncertainty derived from the
power-law fit and from the scatter of the cleaned backbone about that fit. Two
families of outputs are distinguished. Quantities averaged over the fitted lane
(the mean frequency, the mean drift rate, and the mean shock height and speed
derived from them) inherit their uncertainty from the dispersion of the
corresponding per-sample values along the fit. Quantities evaluated at the burst
onset (the initial shock height and speed) are instead governed by the starting
frequency and by the residual scatter of the backbone about the fitted curve.

\subsubsection{Fit-parameter standard errors}
\label{sec:unc-fit}
The power-law model $f(t)=A\,t^{b}$ (Equation~\ref{eqn:power_law}) is fitted to the
cleaned backbone with \texttt{scipy.optimize.curve\_fit} \citep{virtanen2020scipy},
which returns the optimal parameters $(\hat{A},\hat{b})$ together with the
$2\times2$ covariance matrix $\boldsymbol{\Sigma}$. The standard errors of the two
parameters are taken from its diagonal,
\begin{equation}
\sigma_A=\sqrt{\Sigma_{00}}, \qquad \sigma_b=\sqrt{\Sigma_{11}} .
\label{eq:sigmaAb}
\end{equation}
No bootstrap or resampling step is required.

\subsubsection{Instantaneous drift rate}
\label{sec:unc-drift}
The instantaneous drift follows from the fitted derivative,
$g(t)=A\,b\,t^{\,b-1}$ (Equation~\ref{eqn:inst_drift}). Its fractional uncertainty at
each time sample is obtained by combining the fractional errors of the two fit
parameters in quadrature,
\begin{equation}
\sigma_{g}(t)=\lvert g(t)\rvert\,
\sqrt{\left(\frac{\sigma_A}{A}\right)^{2}+\left(\frac{\sigma_b}{b}\right)^{2}} .
\label{eq:sigmadrift}
\end{equation}
This per-sample uncertainty is propagated into the onset shock parameters
described in Section~\ref{sec:unc-onset}.

\subsubsection{Averaged quantities}
\label{sec:unc-avg}
The frequency, drift rate, shock height, and shock speed are evaluated on a dense
set of $N$ samples spanning the fitted lane. For each quantity $X$ the reported
average is the sample mean $\bar{X}=N^{-1}\sum_{j} X_j$, and the associated
uncertainty is the standard error of that mean,
\begin{equation}
\sigma_{\bar{X}}=\frac{1}{\sqrt{N}}
\left[\frac{1}{N}\sum_{j=1}^{N}\left(X_j-\bar{X}\right)^{2}\right]^{1/2} .
\label{eq:sem}
\end{equation}
This is applied uniformly to the mean drift rate $\bar{D}$, the mean frequency
$\bar{f}$, the mean shock height $\bar{R}_s$, and the mean shock speed
$\bar{V}_s$, with $X_j$ the value of the relevant quantity at the $j$-th sample.

\subsubsection{Onset quantities}
\label{sec:unc-onset}
The starting frequency $f_s$ is taken as the upper-percentile ($90$th percentile)
value of the fitted frequency track, representing the high-frequency onset of the
burst. Its uncertainty is estimated from the root-mean-square scatter of the
cleaned backbone about the fitted curve,
\begin{equation}
\sigma_{f}=\left[\frac{1}{N_d}\sum_{k=1}^{N_d}
\left(f_k-\hat{f}_k\right)^{2}\right]^{1/2},
\label{eq:sigmaf}
\end{equation}
where $f_k$ are the backbone frequencies, $\hat{f}_k$ the corresponding fitted
values, and $N_d$ the number of backbone samples. Writing
$\Lambda(f)=\ln\!\left(f^{2}/n_0\kappa^{2}\right)$, the shock height
(Equation~\ref{eqn:shock_height}) and speed (Equation~\ref{eqn:shock_speed}) reduce to
$R_s(f)=\alpha R_\odot\ln10/\Lambda$ and $V_s=\kappa_V(f)\,D$, with sensitivity
coefficients
\begin{equation}
\kappa_R(f)=\frac{2\,\alpha\ln10}{f\,\Lambda^{2}},\qquad
\kappa_V(f)=R_\odot\,\kappa_R(f).
\label{eq:kappas}
\end{equation}
The onset shock height and speed and their uncertainties are then
\begin{equation}
\sigma_{R_s}=\kappa_R(f_s)\,\sigma_f,\qquad
\sigma_{V_s}=\kappa_V(f_s)\,\sigma_g(f_s),
\label{eq:onseterr}
\end{equation}
where $\sigma_g(f_s)$ is the per-sample drift uncertainty (Equation~\ref{eq:sigmadrift})
evaluated at the onset. When the harmonic lane is analyzed, the plasma frequency
and its drift are first mapped through $f_{\rm pe}=f_{\rm H}/2$ and
$\mathrm{d}f_{\rm pe}/\mathrm{d}t=\tfrac{1}{2}\,\mathrm{d}f_{\rm H}/\mathrm{d}t$,
and the corresponding uncertainties are scaled by the same factor before the
shock parameters are formed.

\subsection{Example application of the tool}
\label{subsec:example_usage}
A typical analysis begins by loading a metric e-CALLISTO FITS observation, either from local storage or by downloading the relevant files from the archive and importing them into the main window. The example shown here uses the Type II solar radio burst observed by the CALLISTO instrument at Arecibo Observatory on 2 March 2022 (Figure \ref{fig:mainwindow}). The dynamic spectrum is first inspected in its raw form to verify the time interval and frequency coverage (Figure~\ref{fig:raw}). The user then applies background subtraction and intensity clipping to reduce persistent spectral structure and enhance burst contrast (Figure~\ref{fig:bgs}). Once the Type II emission is visible, the lasso tool is used to isolate the relevant lane region. In this example, the fundamental band was selected because it was sufficiently clear and continuously traceable over the analyzed interval (Figure~\ref{fig:IB}).

The isolated spectrum is then converted into a fitted curve by extracting the maximum-intensity frequency at each time sample within the selected region. The resulting $f_{\max}(t)$ curve is displayed in the maximum-intensity window, where outliers can be removed interactively by selecting and deleting spurious points. With the cleaned maximum-intensity backbone, the analyzer performs a power-law fit and evaluates the derivative of the fitted function to obtain the drift-rate profile and the average drift rate. The shock source height and shock speed are then computed using the one-fold Newkirk density model and the measured plasma-frequency drift (Figure~\ref{fig:analyzer}).

For the Arecibo event, the fitted fundamental-band backbone gave an average drift rate of
\begin{equation}
D = -0.0400 \pm 0.0003~{\rm MHz~s^{-1}}.
\end{equation}
Using the one-fold Newkirk density model, the corresponding average shock height was
\begin{equation}
R_s = 1.715 \pm 0.002~R_{\odot},
\end{equation}
and the average shock speed was
\begin{equation}
V_s = 449 \pm 1~{\rm km~s^{-1}}.
\end{equation}
These values demonstrate the complete analysis path from a raw e-CALLISTO FITS dynamic spectrum to physically interpretable Type II burst parameters. They also show that the software output is not limited to visualization, but can be used to derive quantitative shock diagnostics from a user-verified burst lane.

The quoted uncertainties were obtained using the procedure described in Section~\ref{sec:uncertainty}: the average drift, height, and speed are reported with the standard error of the mean of their per-sample values along the fitted lane, while the onset shock height and speed are propagated from the residual scatter of the backbone about the fit and from the fitted drift-rate uncertainty.

The numerical outputs from this example are summarized in Table~\ref{tab:arecibo_results}. The same result values are also exported by the analyzer as tabulated output, allowing the measurement to be reproduced, compared with other events, or included in later statistical studies.

\begin{table}
\centering
\caption{Example Type II burst parameters obtained from the Arecibo Observatory e-CALLISTO observation on 2 March 2022 using the isolated fundamental band and the one-fold Newkirk density model.}
\label{tab:arecibo_results}
\footnotesize
\setlength{\tabcolsep}{4pt}
\renewcommand{\arraystretch}{1.15}
\begin{tabular}{|p{0.44\columnwidth}|p{0.42\columnwidth}|}
\hline
\multicolumn{1}{|c|}{\textbf{Parameter}} &
\multicolumn{1}{c|}{\textbf{Value}} \\
\hline
\hline
Selected emission lane & Fundamental band \\
\hline
Density model & One-fold Newkirk model \\
\hline
Average drift rate & $-0.0400 \pm 0.0003~{\rm MHz~s^{-1}}$ \\
\hline
Average shock height & $1.715 \pm 0.002~R_{\odot}$ \\
\hline
Average shock speed & $449 \pm 1~{\rm km~s^{-1}}$ \\
\hline
\end{tabular}
\end{table}

\subsection{Fundamental versus harmonic lane comparison}
\label{sec:harmonic-comparison}

\begin{figure}
\centering
\includegraphics[width=\columnwidth]{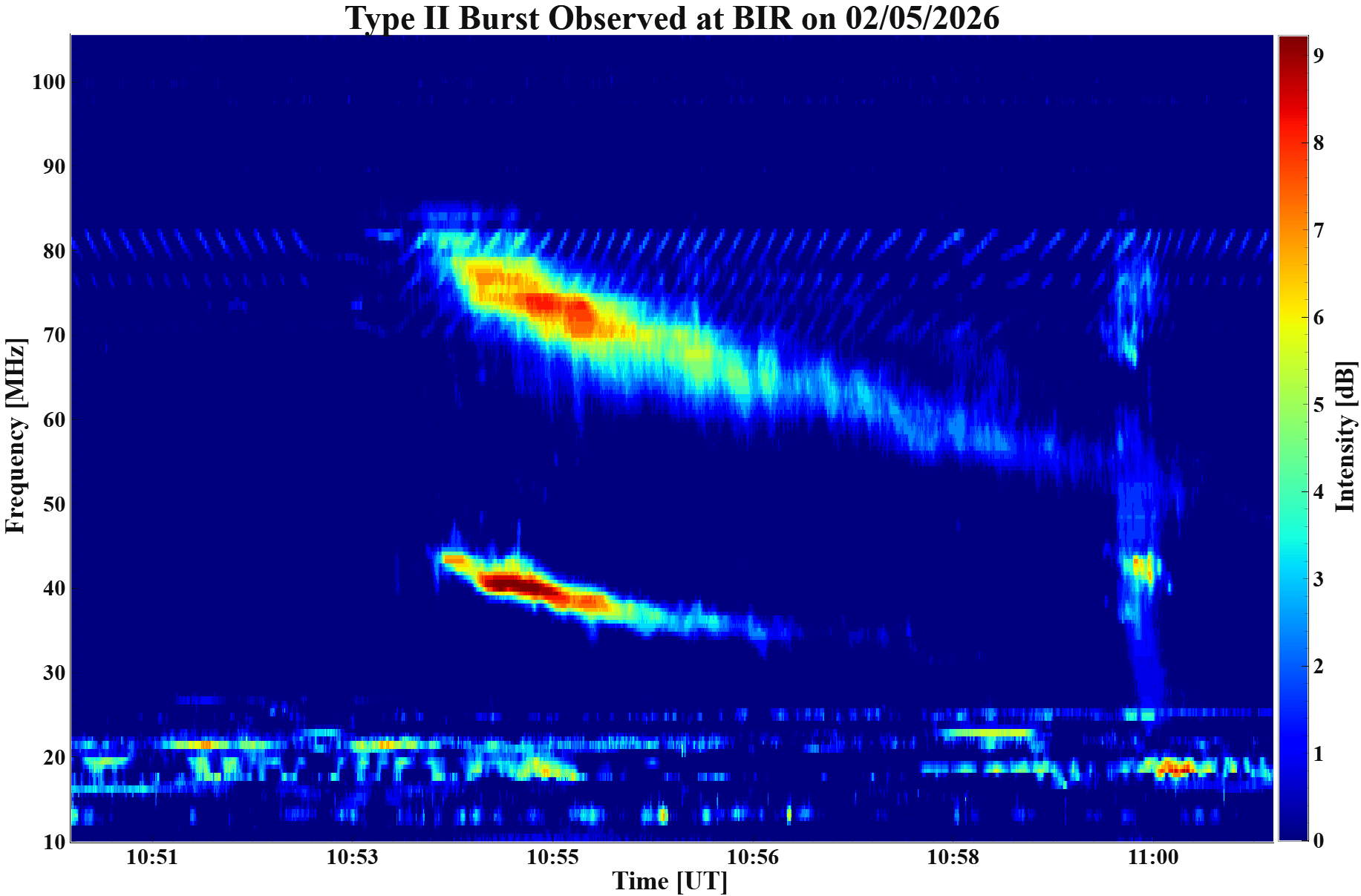}
\caption{Background-subtracted dynamic spectrum of the Type~II solar radio burst
recorded by the CALLISTO instrument at Birr, Ireland on 2~May~2026, showing the
clearly traceable fundamental and harmonic bands used for the independent
lane comparison summarised in Table~\ref{tab:harmonic-comparison}.}
\label{fig:bir-burst}
\end{figure}

Section~\ref{subsec:shock_parameters} argued on analytic grounds that, because the shock speed depends on
the burst only through the logarithmic drift $\mathrm{d}\ln f_{\rm pe}/\mathrm{d}t$,
selecting the harmonic lane in place of the fundamental should perturb the
drift-derived shock speed only weakly and the shock height only through the
logarithm of frequency. To test this directly, we analyzed a Type~II burst
recorded by the CALLISTO instrument at Birr, Ireland on 2~May~2026, an event in
which the fundamental and harmonic lanes are both clearly traceable over the same
interval (Figure~\ref{fig:bir-burst}). Each lane was isolated and processed
independently using the identical workflow: the fundamental frequency was treated
directly as the plasma frequency, while the harmonic backbone was fitted on its
own and converted through $f_{\rm pe}=f_{\rm H}/2$ before the shock parameters
were evaluated. The two lanes therefore provide fully independent estimates of
the same physical quantities, with no parameter carried over from one to the
other. The implied harmonic ratio for this event,
$r = 2\,\bar{f}_{\rm pe}^{\rm (H)}/\bar{f}_{\rm pe}^{\rm (F)} \approx 1.91$, lies
within the $r = 1.9$--$2.1$ range considered in Section~4.3.

\begin{table*}
\caption{Comparison of Type~II shock parameters derived independently from the
fundamental and harmonic lanes of the burst recorded at Birr, Ireland on
2~May~2026, using the one-fold Newkirk density model. The fractional difference
is $|X_{\rm F}-X_{\rm H}|/|X_{\rm F}|\times100\%$, taken relative to the
fundamental-band value.}
\label{tab:harmonic-comparison}
\begin{tabular}{@{}lccc@{}}
\toprule
Parameter & Fundamental & Harmonic & Difference (\%) \\
\midrule
Average frequency (MHz)            & $34.55 \pm 0.10$     & $33.03 \pm 0.10$     & 4.4 \\
Average drift rate (MHz\,s$^{-1}$) & $-0.0426 \pm 0.0005$ & $-0.0414 \pm 0.0005$ & 2.8 \\
Starting frequency (MHz)           & $39.33 \pm 0.93$     & $38.34 \pm 0.58$     & 2.5 \\
Initial shock speed (km\,s$^{-1}$) & $528 \pm 25$         & $554 \pm 14$         & 4.9 \\
Initial shock height ($R_\odot$)   & $1.624 \pm 0.013$    & $1.638 \pm 0.008$    & 0.9 \\
Average shock speed (km\,s$^{-1}$) & $490 \pm 1$          & $512 \pm 1$          & 4.5 \\
Average shock height ($R_\odot$)   & $1.700 \pm 0.003$    & $1.728 \pm 0.003$    & 1.6 \\
\bottomrule
\end{tabular}
\end{table*}

Table~\ref{tab:harmonic-comparison} lists the resulting parameters together with
their fractional differences, defined as
$|X_{\rm F}-X_{\rm H}|/|X_{\rm F}|\times100\%$ relative to the fundamental value.
All seven quantities agree to better than $5\%$, and the pattern of the residuals
follows the expectation of Section~4.3. The shock heights, which depend on
frequency only through $\Lambda = \ln\!\left(f^{2}/n_{0}\kappa^{2}\right)$,
differ by just $1.6\%$ (average) and $0.9\%$ (onset), within the $\lesssim1.7\%$
bound quoted for a $\pm5\%$ departure of the harmonic ratio from $2$. The shock
speeds differ by $4.5\%$ (average) and $4.9\%$ (onset); these are modestly larger
than the $\lesssim3.5\%$ obtained in Section~\ref{subsec:shock_parameters} for a pure rescaling of the
frequency axis, because the two lanes are fitted independently and their
logarithmic drifts are not constrained to coincide, so the measured drifts
contribute a small additional scatter on top of the weak $\Lambda$ dependence.
The differences nonetheless remain at the few-percent level and are comparable
to the statistical uncertainties of the individual fits.

These results confirm empirically that the harmonic lane is a reliable fallback
for shock-parameter estimation when the fundamental lane is weak, fragmented, or
obscured. Under the one-fold Newkirk model the bias introduced by analyzing the
harmonic instead of the fundamental remains below $2\%$ in shock height and of
order $5\%$ in shock speed, in agreement with the analytic invariance argument of
Section~\ref{subsec:shock_parameters}. The fundamental band is therefore retained as the default whenever
it is sufficiently clear, with the harmonic branch used only when the fundamental
is not reliably traceable.

\section{Conclusions}

This work introduces the e-CALLISTO FITS Analyzer as a practical, reproducible
framework for processing and analyzing e-CALLISTO FITS dynamic spectra in a
single GUI environment. Relative to existing archive utilities, scripted
libraries such as \texttt{pyCallisto}, and automated or semi-automated
burst-detection tools (Section~\ref{subsec:novelty}), the contribution of this work is to
combine archive-based retrieval, FITS parsing, time- and frequency-axis
merging, RFI mitigation, interactive burst isolation, and Type~II
shock-parameter estimation into a single transparent, user-verified workflow.
The application addresses common bottlenecks in solar radio burst studies by
streamlining FITS ingestion, supporting archive download with preview, and
providing robust mechanisms to merge fragmented observations across time and
frequency into a continuous spectrum suitable for event-level analysis.

The core processing workflow combines a deterministic, statistics-based
cleaning pipeline with interactive control. Narrow-band interference and
persistently contaminated channels are identified through a two-dimensional
median filter and a robust per-channel $z$-score, repaired from neighbouring
frequency rows, and clipped on a row-by-row percentile basis; row-wise
background subtraction and optional noise equalization then produce a
cleaned, reproducible spectrum. Threshold clipping allows the user to tune
contrast without reloading data, and polygon-based masking enables burst
isolation without imposing a rigid burst geometry. Intensity is reported
throughout in instrumental or relative digit-to-dB units; the present release
does not perform absolute radiometric calibration into solar flux units, and
this is treated explicitly as a stated boundary of the methodology rather
than an implicit assumption.

For Type~II bursts, the analyzer extends this interactive approach into
quantitative measurement. A maximum-intensity backbone is extracted from the
isolated lane, refined through point-wise outlier removal, and fitted with a
power-law model to obtain the frequency drift rate, which is converted into
shock height and shock speed under a selectable coronal density model. The
fundamental band is used whenever it is sufficiently clear; when the harmonic
lane is selected instead, it is fitted independently and converted to the
equivalent plasma frequency, and the logarithmic-drift invariance of the
shock-speed expression (Section~\ref{subsec:shock_parameters}) shows that this conversion leaves the
drift-derived part of the result essentially unaffected, with departures from the nominal harmonic ratio contributing only a small, quantified residual sensitivity. This expectation is confirmed empirically in Section~\ref{sec:harmonic-comparison}, where independently analyzing the fundamental and harmonic lanes of a burst in which both are clearly visible (Figure~\ref{fig:bir-burst}) yields shock heights and speeds that agree to within about $5\%$ (Table~\ref{tab:harmonic-comparison}), validating the harmonic lane as a reliable fallback when the fundamental is weak, fragmented, or obscured. Every reported drift rate, shock height, and shock speed carries an uncertainty derived from the power-law fit: the fit-parameter standard errors set the per-sample drift uncertainty, the averaged quantities are reported with the standard error of the mean of their per-sample values, and the onset quantities are propagated from the residual scatter of the backbone about the fit (Section~\ref{sec:uncertainty}).

The software is designed to remain extensible. In addition to the radio
pipeline, space-weather context modules are integrated as separate
components, including GOES X-ray flux visualization and SOHO/LASCO CME
browsing, enabling event inspection within the same application ecosystem
while keeping the radio workflow stable.

Several limitations remain and also guide future development. Time and
frequency merging depend on compatibility assumptions and do not
automatically correct gaps, overlaps, or station timing irregularities.
Background subtraction, while effective for many cases, can suppress weak
diffuse emission when the background varies rapidly, and burst isolation
still depends on user-defined regions. The current release reports cleaned
spectra in instrumental or relative dB units rather than absolute solar flux
units, since station-specific antenna gain, receiver response, and
system-temperature information are not uniformly available across the
network. Future work will focus on improving robustness for heterogeneous
station configurations, adding more automated assistance for burst candidate
selection while retaining user control, exposing the RFI hot-channel mask and
associated quality flags in exported data products for downstream filtering,
and expanding physical analysis options such as additional density models
and calibration-aware products for studies that require absolute intensity
interpretation.


\section*{Acknowledgements}
We thank the Fachhochschule Nordwestschweiz (FHNW), Institute for Data Science in Brugg/Windisch, Switzerland, for hosting the e-CALLISTO network. The authors also acknowledge the efforts of individual CALLISTO operators, including the Arecibo Observatory, Puerto Rico. The Arecibo Observatory was operated by the University of Central Florida under a cooperative agreement with the National Science Foundation (AST-1822073). We gratefully acknowledge the Department of Physics, University of Colombo, and the Astronomical and Space Science Unit for providing the resources and institutional support that made this work possible. 

\section*{Conflict of Interest}

The authors declare that there are no financial or non-financial conflicts of interest associated with this research, its authorship, or its publication.
\section*{Data Availability}

The solar radio data used in this work are provided by the e-CALLISTO network and are publicly available at the official archive (\url{https://www.e-callisto.org/Data/data.html}). The source code for the \textit{e-CALLISTO FITS Analyzer} is open-source and can be accessed via GitHub at \url{https://github.com/SaanDev/e-Callisto_FITS_Analyzer}. Any additional derived data products or specific software environment manifests are available from the corresponding author upon reasonable request.




\bibliographystyle{rasti}
\bibliography{reference} 



\bsp	
\label{lastpage}
\end{document}